# Electroweak instantons at non-zero Weinberg angle[*]

**M.J. Gibbs**[†],
Cavendish Laboratory, University of Cambridge
Madingley Road, Cambridge CB3 0HE, U.K.

## Abstract

Previous calculations of instanton effects in electroweak theory have concentrated on the case of zero Weinberg angle $\theta_w$, where the U(1) hypercharge field decouples. In this paper we extend the instanton calculation to non-zero $\theta_w$, by constructing a perturbation expansion. This allows for the first time the study of photon production at B and L number violating verticies. We find that the orientation of the instanton solution in isospin space has to be carefully considered to avoid unphysical results.

Cavendish–HEP–94/5
January 10, 1995

---

[*]Research supported in part by the UK Particle Physics and Astronomy Research Council and by the EC Programme "Human Capital and Mobility", Network "Physics at High Energy Colliders", contract CHRX-CT93-0357 (DG 12 COMA).
[†]Current address: Department of Physics, Oliver Lodge Laboratory, Liverpool University, U.K.

# 1 Introduction

It has been known for some time that topologically non-trivial gauge fields (*instantons*) induce baryon and lepton number violating verticies in electroweak theory [1, 2]. More recent investigations into these processes suggest that such processes may be observable at supercollider energies. An extensive review can be found in [3].

However, present calculations have all been in the context of SU(2)-Higgs models, which correspond to the limit $\sin^2 \theta_w = 0$. In this case the U(1)$_Y$ field decouples and can be consistently set to zero. In order to investigate more fully the possibility of observing B and L violating processes at supercollider energies [4], one has to include the effect of a non-zero Weinberg angle. This will allow the study of B and L violating vertices which include photons as well as massive gauge bosons for the first time.

In this paper we consider the inclusion of the U(1)$_Y$ field by a perturbative expansion in $\theta_w$, similar to the method discussed in [5–7] for the sphaleron. We construct the instanton which dominates the euclidian path integral in the unit winding number sector of the electroweak theory, and calculate the effect of non-zero $\theta_w$. The resultant expressions appear to violate co-ordinate system independence, and we demonstrate how this problem is resolved by considering the integration over the instanton orientations.

Our discussion is limited to the bosonic part of the electroweak theory, in euclidian space. This is obtained by performing a Wick rotation $x_0 \to -ix_0$. The corresponding rotation for gauge fields is $A_0 \to iA_0$. The relevant part of the Lagrangian is then

$$\mathcal{L}_{boson} = \frac{1}{2}\text{Tr}\left(F_{\mu\nu}F_{\mu\nu}\right) + \frac{1}{4}\left(B_{\mu\nu}B_{\mu\nu}\right) + \left(D_\mu\phi\right)^\dagger D_\mu\phi + \lambda\left(\phi^\dagger\phi - \frac{1}{2}v^2\right)^2. \tag{1}$$

The two gauge field tensors are defined in terms of the corresponding gauge fields in the usual way. Thus, for the SU(2) gauge field $A_\mu = \frac{\sigma_a}{2}A_\mu^a$ we have

$$F_{\mu\nu} = \partial_\mu A_\nu - \partial_\nu A_\mu + [A_\mu, A_\nu] \tag{2}$$

and for the U(1) field $B_\mu$

$$B_{\mu\nu} = \partial_\mu B_\nu - \partial_\nu B_\mu. \tag{3}$$

The symmetry is spontaneously broken by the Higgs expectation value $\langle\phi\rangle = v/\sqrt{2}$, leaving U(1)$_{\text{EM}}$. This can be demonstrated by writing the field $\phi$ in the form

$$\phi(x) = \frac{1}{\sqrt{2}}(v + \eta(x))\begin{pmatrix} 0 \\ 1 \end{pmatrix} \tag{4}$$

This choice for the expansion of $\phi$ corresponds to choosing the unitary gauge. The $W^\pm$ bosons correspond to the fields

$$W_\mu^\pm = \frac{1}{\sqrt{2}}\left(A_\mu^1 \mp iA_\mu^2\right) \tag{5}$$

and the $Z^0$ boson and photon can be written in terms of the Weinberg angle $\tan\theta_w = g'/g$ as

$$Z_\mu = \cos\theta_w A_\mu^3 - \sin\theta_w B_\mu \tag{6}$$
$$\gamma_\mu = \sin\theta_w A_\mu^3 + \cos\theta_w B_\mu \tag{7}$$

respectively. The masses of these particles are then $m_\gamma = 0$, $m_W = gv/2$, and $m_Z = m_W/\cos\theta_w$ for the photon and the $W^\pm$ and $Z^0$ bosons respectively. The Higgs boson, which is the physical manifestation of the field $\eta$, has mass $m_H = \sqrt{2\lambda}v$.

In our notation, the classical equations of motion of the fields in (1) are

$$D_\nu F_{\mu\nu} = J_\mu^{\text{SU}(2)_L}, \qquad (8)$$

$$\partial_\nu B_{\mu\nu} = J_\mu^{\text{U}(1)_Y}, \qquad (9)$$

$$D_\mu D_\mu \phi = 2\lambda \left(\phi^\dagger \phi - \frac{1}{2}v^2\right) \phi, \qquad (10)$$

and the associated currents are

$$J_\mu^{\text{SU}(2)_L} = -\frac{1}{4}ig \left(\phi^\dagger \sigma^a D_\mu \phi - (D_\mu \phi)^\dagger \sigma^a \phi\right) \qquad (11)$$

$$J_\mu^{\text{U}(1)_Y} = -\frac{1}{2}ig' \left(\phi^\dagger D_\mu \phi - (D_\mu \phi)^\dagger \phi\right). \qquad (12)$$

Introduction of fermion fields into the calculation of S-matrix amplitudes introduces a coupling between the determinant of the quantum fluctuations about the classical instanton solution and the fermionic fields in such a way as to violate B and L number conservation. The contribution to the determinant of this coupling is referred to as the 'zero-modes' of the fermionic fields.

## 2 Instantons

The existence of non-trivial solutions to the field equations of SU(2) theory has been known for some time [8]. For such a theory, with no Higgs or U(1) gauge field, the classical equation of motion is $D_\mu F_{\mu\nu} = 0$. The solution with unit winding number is referred to as the *instanton*. In the singular gauge the expression for an instanton located at the origin is

$$A_\mu^a(x) = \frac{2}{g}\frac{\rho^2}{x^2}\frac{1}{x^2 + \rho^2}\overline{\eta}_{\mu\nu}^a x_\nu, \qquad (13)$$

where the $\overline{\eta}_{\mu\nu}^a$ are the 't Hooft symbols [1]. The parameter $\rho$ is the size of the instanton. The corresponding expression for an instanton centred at $z$ is obtained by using $x_\mu \to (x-z)_\mu$.

However, a complication arises when the Higgs field is included. It can be shown that in this case the equations have no nontrivial solutions with finite action by using simple scaling arguments [9]. This makes the semicalssical evaluation of the euclidian path integral by an expansion about nontrivial configurations difficult [10, 11]. The constrained instanton [9] of Affleck is a suitable method for dealing with this problem. We now briefly outline the construction of the constrained instanton for the case $\theta_w = 0$ before proceeding the the general case of non-zero $\theta_w$.

A constraint is introduced into the path integral

$$1 = \int d\rho \Delta(\rho) \, \delta \left(\int d^4x \, \mathcal{O} - \rho^{d-4}\right) \qquad (14)$$

where $\mathcal{O}$ is a local operator of dimension $d > 4$ and $\Delta$ is a Jacobian. Fourier transforming this $\delta$-function allows the path integral to be performed with $\rho$ fixed. This is equivalent to calculating the path integral with the addition of an extra term $+i\omega\mathcal{O}$ into the Lagrangian, where $\omega$ is the conjugate Fourier variable to $\rho$. Both the path integral and the $\omega$ integral now have a saddle point, and can be performed by the method of steepest descents. Operators such as $\text{Tr}\, F^3$ and $(\phi^\dagger \phi - v^2/2)^3$ yield suitable classical solutions. They will also have no effect to lowest order in perturbation theory on the B and L number violating fermion zero modes as they are constructed only of boson fields.

The classical solution obtained by this procedure is the constrained instanton. The centre of the solution, at small $r = |x|$, is a 'core' where the unconstrained instanton and the associated Higgs solution are approximate solutions. In the framework of the constrained instanton approach these solutions are valid in the limit $r \ll \rho \ll 1/v$, where the source terms in (8,9,10) can be neglected. Outside this core, at large distances such that $r \gg \rho$, the source terms become important. In this limit the fields are dominated by the solutions of

$$A_\mu^{acl}(x) = -\frac{P_W(\rho)}{g} \overline{\eta}_{a\mu\nu} \partial_\nu G_{m_W}(x) \qquad (15)$$

$$\phi^{cl}(x) = \frac{v}{\sqrt{2}} [1 - P_H(\rho) G_{m_H}(x)] \begin{pmatrix} 0 \\ 1 \end{pmatrix}, \qquad (16)$$

where $G_m(x)$ satisfies

$$\left(-\partial^2 + m^2\right) G_m(x) = \delta^{(4)}(x). \qquad (17)$$

Using the solution

$$G_m(x) = \frac{1}{4\pi^2} \frac{m}{x} K_1(mx), \qquad (18)$$

the functions $P_W(\rho)$ and $P_H(\rho)$ are found by matching the two solutions in the intermediate region where $r \sim \rho$. By working to leading order in $\rho$, the operator $\mathcal{O}$ used to constrain the instanton in Eq. (14) can be neglected, and we find $P_W(\rho) = 4\pi^2 \rho^2$ and $P_H(\rho) = 2\pi^2 \rho^2$. This large $x$ behaviour, after Fourier transforming, becomes an effect at small $k$. We find that poles of Green functions based on expansions about the constrained instanton [10,11] are at $m_W^2$ and $m_H^2$ for the gauge and Higgs bosons respectively. Note that a computation based on the unconstrained instanton would yield poles for $k^2 \to 0$ for all boson types, corresponding to $m_W = m_H = 0$.

## 2.1 Electroweak symmetry breaking

The discussion of the constrained instanton in the literature has so far been in the context of SU(2) gauge theories only, which corresponds to $\sin\theta_w = 0$ in the standard model. We now extend this discussion to include the study of instanton processes for non-zero $\theta_w$. Our approach is to introduce the symmetry breaking by forming a perturbation expansion in $\theta_w$.

The U(1)$_Y$ current (12) will provide a source for the $B_\mu$ field. The leading order correction in $\theta_w$ can be computed by the following procedure. We assume that the scalar and SU(2)$_L$ fields are unchanged. The U(1)$_Y$ current is calculated using (12) with the $B_\mu$ contribution to the covariant derivative of $\phi$ held at zero, and finally equation (9) is solved to yield the induced U(1)$_Y$ gauge field. Higher order corrections can then be determined by computing the changes in the other fields and iterating the procedure.

The general form of the lowest-order background fields from the constrained instanton computation are

$$A_\mu^{a\,cl}(x) = \frac{1}{g}\overline{\eta}_{a\mu\nu}\, x_\nu\, \alpha(x;\rho), \qquad (19)$$

$$\phi^{cl}(x) = \frac{v}{\sqrt{2}}\,\beta(x;\rho)\begin{pmatrix}0\\1\end{pmatrix}. \qquad (20)$$

Inserting this form into the field equations and using the procedure just described, we find the U(1)$_Y$ current to be

$$J_\mu^{U(1)_Y}(x) = g'\frac{v^2}{4}\beta^2(x;\rho)\overline{\eta}_{3\mu\nu}\,x_\nu\,\alpha(x;\rho). \qquad (21)$$

We now impose the Landau gauge, $\partial_\mu B_\mu = 0$. In this gauge

$$-\partial^2 B_\mu = J_\mu^{U(1)_Y} \qquad (22)$$

and the Fourier transform $\bar{B}_\mu^{cl}(k)$ of the induced classical field is then

$$\bar{B}_\mu^{cl}(k) = ig'v^2\pi^2\overline{\eta}_{3\mu\nu}\frac{k_\nu}{k^4}\int_0^\infty \mathrm{d}r\, r^3 J_2(kr)\,\beta^2(r)\,\alpha(r). \qquad (23)$$

Near the mass shell,[‡] $k^2 + m^2 \to 0$, the large distance behaviour of the classical fields dominates the Fourier transform. Therefore, we insert the large distance form of the constrained instanton for the functions $\alpha$ and $\beta$ into Eqs. (19,20,23). For the Higgs field we find

$$\bar{\eta}^{cl}(k) = -\frac{v}{\sqrt{2}}\frac{2\pi^2\rho^2}{k^2+m_H^2}, \quad \text{for } k^2+m_H^2 \to 0. \qquad (24)$$

The $W^\pm$ fields are

$$\bar{W}_\mu^{\pm\,cl}(k) = i\frac{4\pi^2}{g}\rho^2\left[\overline{\eta}_{1\mu\nu} \mp i\overline{\eta}_{2\mu\nu}\right]\frac{k_\nu}{k^2+m_W^2}, \quad \text{for } k^2+m_W^2 \to 0, \qquad (25)$$

and for the $Z^0$ boson and photon we obtain

$$\bar{Z}_\mu^{0\,cl}(k) = \cos\theta_w\, i\frac{4\pi^2}{g}\rho^2\overline{\eta}_{3\mu\nu}\frac{k_\nu}{k^2+m_Z^2}, \quad \text{for } k^2+m_Z^2 \to 0, \qquad (26)$$

$$\bar{\gamma}_\mu^{cl}(k) = \sin\theta_w\, i\frac{4\pi^2}{g}\rho^2\overline{\eta}_{3\mu\nu}\frac{k_\nu}{k^2}, \quad \text{for } k^2 \to 0. \qquad (27)$$

Two comments about these last two equations are in order. Firstly, the leading coefficients of $\cos\theta_w$ and $\sin\theta_w$ arise from the superposition of the $A_\mu^3$ and $B_\mu$ fields according to Eqs. (6,7), and are separate to factors of $\theta_w$ arising from our perturbation expansion. Secondly, the shift of the position of the poles of the classical field is exact only in the case of the photon, Eq. (27). Strictly speaking, the $Z^0$ boson pole is still at $m_W^2$. This is because we are working only to leading order in $\theta_w$. Therefore the difference between $m_W$ and $m_Z$ is considered to be small - they are equal to $\mathcal{O}(\theta_w)$.

---

[‡] Recall that we are working in Euclidian space-time.

A straightforward interpretation of Eqs. (26) and (27) can be made. The production of the $Z^0$ boson and the photon from the instanton can be pictured as a two stage process: the emission of a $W^0$ boson, followed by the decay $W^0 \to Z^0$ or $W^0 \to \gamma$. Construction of the Feynman diagrams for these decays, as shown in Fig. (1) leads to the extra factors

$$\cos\theta_w \; \frac{k^2 + m_W^2}{k^2 + m_Z^2} \quad \left(W^0 \to Z^0\right) \tag{28}$$

and

$$\sin\theta_w \; \frac{k^2 + m_W^2}{k^2} \quad \left(W^0 \to \gamma\right) \tag{29}$$

which replace the $W^0$ mass pole with a pole at the relevant physical particle mass. In this picture an emitted $W^0$ decays into a $Z^0$ boson or a photon with relative probabilities $\cos^2\theta_w$ and $\sin^2\theta_w$ respectively. Treating physical particle production as the decay products of the $W^0$ boson is identical to inserting the expressions (26) and (27) into the constrained instanton calculation.

## 2.2 Integration over the instanton orientation

The 't Hooft symbols in the expressions we have derived for the classical fields couple weak isospin and space-time indices. Insertion of Eqs. (25) to (27) into Green functions would therefore imply that the breaking of the $SU(2)_L \times U(1)_Y$ symmetry produces instanton effects which are coordinate system dependent. However, the instanton solutions we have discussed can be rotated in $SU(2)$ space. The effect of the rotation is to change the relative orientation of isospin space to real space of the classical solution. The fermion zero modes are $SU(2)$ doublet eigenfunctions of operators formed from the classical instanton field, and so are also rotated in $SU(2)$ space. In order to obtain the correct, coordinate system independent result, we have to sum over all possible orientations. Note that the orientation of the $\phi$ field is uncorrelated to the gauge field orientation, as the action is invariant under *independent* rotations of the $A_\mu$ and $\phi$ fields [11]. The orientation of the $\phi$ field is completely fixed by spontaneous symmetry breaking.

This summation corresponds to integrating over all possible rotation matrices in the $SU(2)$ group. As well as removing the coordinate system dependence, this procedure has two other effects. One of these is to enforce charge conservation. Particle configurations that violate conservation of charge have contributions which are finite for a particular orientation, but upon integration give no contribution to the matrix elements. The second effect is to produce a large number of terms in the expression for the matrix element. The origin of these terms can be seen by studying the integration of a product of $SU(2)$ rotation matrices over all possible rotations.

A general SU(2) rotation matrix $U$ can be written in the form $U = n_\mu \sigma_\mu$, where the four vector $\sigma_\mu \equiv (i, \sigma)$. The integral of two rotation matrices, both corresponding to the same rotation but acting on different isospin vectors $|a\rangle$ and $|b\rangle$, is

$$\int_{\mathrm{SU}(2)} dU \ U|a\rangle \ U|b\rangle = \sigma_\mu |a\rangle \ \sigma_\nu |b\rangle \int dn n_\mu n_\nu$$
$$= \sigma_\mu |a\rangle \ \sigma_\nu |b\rangle \ \delta_{\mu\nu}. \quad (30)$$

Here $dU$ is the SU(2) group measure. The numeric constant from the integration is ignored. The integral of an odd number of rotation matrices is exactly zero, so we next consider the case of four rotation matrices. Extending our notation in an obvious way, the result of the integration in this case is

$$\sigma_\mu |a\rangle \ \sigma_\nu |b\rangle \ \sigma_\rho |c\rangle \ \sigma_\sigma |d\rangle \times \left(\delta_{\mu\nu}\delta_{\rho\sigma} + \delta_{\mu\rho}\delta_{\nu\sigma} + \delta_{\mu\sigma}\delta_{\nu\rho}\right), \quad (31)$$

and the generalisation for $2n$ matricies is obvious.

The effect of the integration is to couple the isospin vectors together in a symmetric manner. The integral of $N_R$ rotation matrices (for even $N_R$) will give $(N_R - 1)!!$ terms[§]. We can now see what effect this will have on the structure of the matrix element. The SU(2) rotated classical field $A_\mu^{a\ cl}(U)$ can be expressed in terms of the unrotated field as

$$A_\mu^{a\ cl}(U) = \mathrm{Tr}\left[\sigma^a U \sigma^b U^\dagger\right] A_\mu^{b\ cl}. \quad (32)$$

The effect of the rotation matrix on the fermion zero modes will be to rotate their direction in isospin space, for example $\psi^{cl}(U;x) = U\psi^{cl}(x)$. Hence there will be one rotation matrix in the integral for each of these fields. The integration over all rotations will cause these fields to couple together. This coupling will link the momentum and polarization vectors together in a nontrivial way when the classical solutions are inserted to obtain the matrix element. For the Standard Model, we have $4n_G$ fermion zero modes and there will be $N_R = 4n_G + 2n_B$ rotation matrices.

The total charge of the classical fields is obtained by integrating the time component of the appropriate current. Therefore, the SU(2) charges are given by

$$Q^{a\ cl}(t) = \int d^3x \ J_0^{a\ \mathrm{SU}(2)_\mathrm{L}}(t,\mathbf{x}), \quad (33)$$

and the U(1) charge by

$$Q^{cl}(t) = \int d^3x \ J_0^{\mathrm{U}(1)_\mathrm{Y}}(t,\mathbf{x}). \quad (34)$$

However, we have to take into account the integration over orientations by substitution of Eq. (32) into these expressions and integrating over all orientations. Denoting by $\overline{Q}^a$ the SU(2) charge integrated over all configurations, it is straightforward to show that

$$\overline{Q}^{a\ cl}(t) = Q^{b\ cl}(t) \int dU \mathrm{Tr}\left[\sigma^a U \sigma^b U^\dagger\right]. \quad (35)$$

Substitution of Eq. (30) and taking the trace gives $\overline{Q}^{a\ cl}(t) \equiv 0$ independent of $Q^{b\ cl}$. The same result is obtained for the integrated U(1) charge $\overline{Q}^{cl}$ by considering the effect of rotating the instanton solution in this case also. The fact that these charges are identically zero for all values of $t$ conserves SU(2)$_\mathrm{L}$ and U(1)$_\mathrm{Y}$ separately. Therefore, electric charge, which is a superposition of these two charges, is also conserved.

---
[§]Note that $n!! = n(n-2)(n-4)...3.1$ for odd $n$.

## 3 Summary

In this paper we have constructed the electroweak instanton corresponding to non-zero Weinberg angle. This demonstrates for the first time the coupling of the photon to instanton induced B and L number violating verticies, and we have proposed a simple physical interpretation of the symmetry breaking. The resultant expressions for the classical background fields appear to violate co-ordinate system independence. We have shown that this problem can be resolved by integrating over all possible orientations of the classical field configuration. Such an integration, in the case of the electroweak instanton, forces the interactions to respect electric charge conservation.

The integration over orientations leads to an expression for the matrix element of the process which contains a large number of terms. Calculations of B and L number violating matrix elements due to instanton processes indicate a typical boson multiplicity of $n_B \sim 1/\alpha_W$. Taking $n_B = 30$ and $n_G = 3$ gives a typical value of $N_R \sim 70$. However, taking for example $N_R = 50$, direct calculation of $49!! \sim 10^{32}$ terms is not feasible. This problem can be overcome by noting that the large number of outgoing particles, $\sim 40$, means that a good working assumption of the final state is an isotropic distribution in solid angle [4].

We end by considering the implications for instanton-induced processes in QCD [12–14]. The integration over the SU(3) strong interaction group can essentially be separated into two parts; the SU(2) subgroup containing the instanton and the remainder of the group, although one must be careful as there are some subtleties in this separation [15]. For QCD, the relevant particle multiplicites are much lower than the electroweak case, and consequently the integration over instanton configurations will contain much fewer terms. Therefore, exact computation of the integral may be possible in this case.

## Acknowledgements

The author would like to thank A. Ringwald and B.R. Webber for helpful discussions and comments.

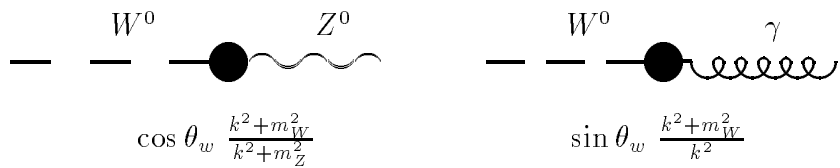

Figure 1: Feynman diagrams for the decays $W^0 \to Z^0$ and $W^0 \to \gamma$.